\documentclass{aa}  
\usepackage{graphicx}
\usepackage{txfonts}
\usepackage{natbib}
\bibpunct{(}{)}{;}{a}{}{,} 

\newcommand \heI{He I~}

\newcommand \hal{H$\alpha$~}
\newcommand \haln{H$\alpha$}
\newcommand \hbeta{H$\beta$~}

\newcommand \rstar{R$_\star$}
\newcommand \kms{km\,s$^{-1}$~}
\newcommand \kmsn{km\,s$^{-1}$}
\begin{document}
   \title{Magnetospheric accretion-ejection processes in the classical
          T Tauri star AA Tau\thanks{Based on observations obtained at
          ESO Obs., Chile, Maidanak Obs., Uzbekistan, Buryakan Obs.,
          Armenia, Konkoly Obs., Hungary, and La Palma Obs., Spain.}}


   \author{J. Bouvier\inst{1}
\and           S.H.P. Alencar\inst{1,2}
\and           T. Boutelier\inst{1}
\and           C. Dougados\inst{1}
\and	   Z. Balog\inst{3,4}
\and	   K. Grankin\inst{5}
\and	   S.T. Hodgkin\inst{6}
\and	   M.A. Ibrahimov\inst{5}
\and       M. Kun\inst{7}
\and	   T.Yu. Magakian\inst{8}
\and       C. Pinte\inst{1}
          }

   \offprints{J. Bouvier}

   \institute{Laboratoire d'Astrophysique de Grenoble, Universit\'e Joseph-Fourier,
              BP 53, 38041 Grenoble Cedex 9, France 
         \and
             Departamento de F\'{\i}sica, ICEx-UFMG, CP. 702,
             Belo Horizonte, MG, 30123-970, Brazil
	     \and Steward Observatory, University of Arizona, 933 N. Cherry Av.,
Tucson AZ 85721 USA
	     \and 
Department of Optics and Quantum Electronics, University of Szeged,
              D\'om t\'er 9, H-6720 Szeged, Hungary
\and
Ulugh Beg Astronomical Institute of the Uzbek Academy of Sciences,
              Astronomicheskaya 33, 700052 Tashkent, Uzbekistan
\and
Institute of Astronomy, Madingley Road, Cambridge CB3 0HA, U.K.
\and Konkoly Observatory, H-1525 Budapest P.O. Box 67
Hungary
\and             
V.A. Ambartsumyan Byurakan Astrophysical Observatory, Armenia
}
   \date{Received ; accepted }

 
\abstract
{Accretion and ejection are complex and related processes that
vary on various timescales in young stars.}
{We intend to investigate the accretion and outflow dynamics and
their interaction from observations of the classical T Tauri star
AA Tau.}
{From a long time series of high resolution (R=115,000)
HARPS spectra and simultaneous broad-band photometry, we report new
evidence for magnetospheric accretion as well as ejection 
processes in the nearly edge-on classical T Tauri star AA Tau.}
{AA Tau's light curve is modulated with a period of 8.22d. The
recurrent luminosity dips are due to the periodic occultation of the
central star by the magnetically-warped inner disk edge located at
about 9 \rstar.  Balmer line profiles exhibit a clear rotational
modulation of high-velocity redshifted absorption components with a
period of 8.22 days as well, with a maximum strength when the main
accretion funnel flow passes through the line of sight. At the same
time, the luminosity of the system decreases by about 1 mag,
indicative of circumstellar absorption of the stellar photosphere by
the magnetically-warped, corotating inner disk edge. The photospheric
and HeI radial velocities also exhibit periodic variations, and the
veiling is modulated by the appearance of the accretion shock at the
bottom of the accretion funnel. Diagnostics of hot winds and their
temporal behaviour are also presented.}
{The peculiar geometry of the young AA Tau system (nearly edge-on)
allows us to uniquely probe the acretion-ejection region close to the
star. We find that most spectral and photometric diagnostics vary as
expected from models of magnetically-channelled accretion in young
stars, with a large scale magnetosphere tilted by 20$\degr$ onto the
star's spin axis. We also find evidence for time variability of the
magnetospheric accretion flow on a timescale of a few rotational
periods. }


\authorrunning{Bouvier et al.} 
    
\titlerunning{Magnetospheric accretion and ejection in AA Tau}

\maketitle
%
\section{Introduction}

For a few million years after their formation, young solar-type stars
(T Tauri stars, TTS) continue to accrete material from their
circumstellar disks. The accretion process has a profound and long
lasting impact on early stellar evolution. It is thought to be the
driving source of the powerful jets originating from young stars and
to regulate their angular momentum before they reach the zero age main
sequence. Accretion is also the central ingredient of disk evolution
at a time when planets start to form, and accretion both into the disk
and onto the star is held responsible for most of the peculiar
properties of T Tauri stars (see, e.g., the review by M\'enard \&
Bertout 1999). Understanding how accretion proceeds into the disk and
onto the star is therefore a major issue for star and planet formation
theories.

It is now commonly acknowledged that accretion from the inner disk
onto the star is, at least in part, mediated by the stellar magnetic
field. Recent Zeeman measurements indicate surface magnetic fields of
a few kilogauss in T Tauri stars (e.g. Johns-Krull et al. 1999, 2004;
Symington et al. 2005a), strong enough to disrupt the inner disk
region and control the accretion flow. It is thus expected that the
inner disk edge lies at a few stellar radii from the central star and
that material is accreted onto the star along funnel flows which
follow the topology of the large scale component of the stellar
magnetosphere. As the free-falling material eventually reaches the
star, the kinetic energy is dissipated in a shock at the stellar
surface (see, e.g. K\"onigl 1991).

Recent numerical simulations of the magnetic star-disk interface
confirm most of these expectations. Assuming a dipolar magnetosphere
on the large scale, whose axis can either be aligned with or inclined
onto the stellar rotation axis, simulations show the development of
the inner disk truncation by strong magnetic fields, leading to a
magnetospheric cavity extending over a few stellar radii
(e.g. Romanova et al. 2002, 2003). In this magnetically-controlled
region, material is accreted onto the star through a complex and time
variable network of funnel flows, whose structure and evolution depend
on the accretion rate, on the topology of the magnetosphere and its
inclination on the stellar axis (e.g. Romanova et al. 2003). As a
result, the accretion shock at the stellar surface can take various
shapes and be located at different latitudes (Romanova et
al. 2004). Both numerical simulations and analytical models suggest
that the inner disk structure is significantly modified by its
interaction with the stellar magnetosphere (Lai 1999, Terquem \&
Papaloizou 2000).

The winds of classical T Tauri stars (CTTSs) have been primarily
investigated through the observations and modelling of forbidden
emission lines, whose high velocity component originates in spatially
extended microjets (Hartigan et al. 1995, Shang et al. 2002).  A few
CTTSs had actually their microjets emission imaged. They arise in a
region very close to the star and were observed with ground based
adaptive optics (Dougados et al. 2000) and HST (Bacciotti et
al. 2002), providing unprecedent information on the dynamics near the
base of the jet, where the wind is formed.  Other high velocity inner
disk wind diagnostics include cooler outflowing gas that absorbs
emission from the accreting material (Edwards 2003) and the observed
hot helium gas in emission (Beristain et al. 2001) as well as P Cygni
profiles with strong blueshifted absorptions reaching well into the
stellar continuum over a broad velocity range (Edwards et al. 2003,
Dupree et al.  2005). The above wind signatures are prominent in CTTSs
with high mass accretion rates, indicating that these are actually
accretion-powered winds.

With an extension of a few stellar radii up to the inner disk edge
($\simeq$ 0.05 AU), the magnetospheric cavity can hardly be resolved
directly, even with the most powerful interferometers (e.g. Akeson et
al. 2005). In addition, the magnetospheric cavity is filled with hot
gas while dust particules are likely sublimated. Hence, diagnotics of
hot plasma would have to be observed at this scale, which requires
interferometric observations at high spectral resolution. For the time
being, only indirect signatures of the structure of the star-disk
interaction region and its physical properties can therefore be
searched for.

As the magnetospheric cavity up to the disk inner edge is thought to
corotate with the central star, a powerful approach to confront model
predictions is to monitor the variations of suited photometric and
spectral diagnostics over a few rotational periods. Synoptic studies
may reveal periodic variability which can then be associated to
physical structures in the rotating magnetosphere and the inner
disk. A number of such studies have been performed in the last decade
providing (sometimes conflicting) clues to the magnetospheric
accretion process in young stars (see Bouvier et al. 2006 for a
review).

Two of our previous monitoring campaigns (Bouvier et al. 1999, 2003,
Paper I and Paper II hereafter) have focused on the classical T Tauri
star AA Tau. Thanks to its high viewing angle ($i \simeq$ 75$\degr$),
the line of sight to the stellar surface intercepts the star-disk
interface close to the disk plane. This fortunate configuration
maximizes the observational signatures of the magnetospheric accretion
process. Furthermore, AA Tau is quite a typical classical T Tauri
star, with a K7 spectral type, moderate accretion rate and Class II
SED. Hence, it is likely that the results obtained for this otherwise
unremarkable CTTS can be extended to the whole class of CTTSs.

From our previous campaigns on AA Tau we reported evidence for the
modulation of photometric and spectroscopic diagnostics with a period
of 8.2 days, the rotational period of the star. The modulation of the
stellar flux was interpreted as resulting from the periodic
occultation of the stellar photosphere by the optically thick,
magnetically-warped inner disk edge located at 8.8 stellar radii
(Bouvier et al. 1999). A second campaign allowed us to relate the
recurrent eclipses of the star by the inner disk to accretion funnel
flows and accretion shocks at the stellar surface (Bouvier et
al. 2003). We also found evidence for large scale changes occurring in
the accretion flow onto the star on a timescale of a few weeks,
presumably resulting from a major disturbance of the underlying
magnetospheric structure.

In this paper, we report the results of a new campaign on AA Tau aimed
at further constraining the structure of the magnetically-dominated
star-disk interaction region. We have monitored the photometric
variations of the system over nearly 5 months in the fall of 2004, and
during part of the photometric campaign, we simultaneously obtained a
series of high resolution (R=115,000) spectra on the object. To our
knowledge, this is the first time such a series of high resolution
spectra have been obtained on a T Tauri star. In Section 2, we
describe the photometric and spectroscopic observations obtained
during this campaign. In Section 3, we present the results on the
photometric and spectroscopic variability of AA Tau, which we find to
be periodic with the same period as reported previously. In Section 4,
we discuss the new evidence provided by these observations for
magnetospheric accretion and ejection processes in this object.

\section{Observations}
\subsection{Photometry}

\begin{table*}[t]
\caption{Journal of photometric observations.}             
\label{phot}      
\centering                          
\begin{tabular}{l l l l l l }        
\hline\hline                 
Observatory
        &JD-2450000.
        &Filters
        &Observer        
        &Country
        &N$_{obs}$(V)
        \\
\hline                        
Maidanak
        &3260.-3311.
        & Tube BV
        &K. Grankin
        &Uzbekistan
        & 19
        \\
Maidanak
        &3307.-3346.
        & CCD BV
        &M. Ibrahimov
        &Uzbekistan
        & 32
        \\
Buryakan
        &3297.-3316
        & CCD BV
        &T. Magakian
        &Armenia
        & 5
        \\
Konkoly
        &3334.-3382
        & CCD BV
        &Z. Balog
        &Hungary
        & 32
        \\
IAC
        &3326.-3396
        & CCD V
        &S. Hodgkin
        &Canarias
        & 28
        \\
\hline                                   
\end{tabular}
\end{table*}

The Journal of photometric observations is given in
Table~\ref{phot}. Observations were carried out from several sites over a
time span of nearly 140 days from September 2004 to January 2005, using
either CCD detectors or photomultiplier tubes. Measurements were obtained
in the B and V filters. Differential photometry was performed on CCD
images and absolute photometry from photomultiplier observations, with an
accuracy of the order of 0.01mag in both filters. Somewhat larger
systematic errors ($\leq$ 0.05 mag) might result from the relative
calibration of the photometry between sites. All data reduction procedures
were previously described in Bouvier et al. (2003).

\subsection{Spectroscopy}
The spectroscopic observations of AA Tau were carried out from October 10
to December 02, 2004, at ESO, La Silla. We obtained 22 high-resolution
spectra, over 22 non consecutive nights spread between JD~3288 and JD~3341
at the 3.6 m telescope with the HARPS dual fiber echelle spectrograph
(Mayor et al. 2003) covering the 3800 \AA \ to 6900 \AA \ spectral domain
at a spectral resolution of $\lambda/\Delta\lambda \approx 115 \ 000$.
Integration times of 1800s yielded a S/N ratio between 10 and 30 at 600 nm
depending on the object brightness.  Quasi-simultaneous photometry was
obtained over the period of the spectroscopic observations (JD~3288-3341,
see Table~\ref{phot}). The Journal of spectroscopic observations is
given in Table~\ref{spec}.

The data was automatically reduced by the HARPS Data Reduction Software.
The reduction procedure includes optimal extraction of the orders and
flat-fielding, achieved through a tungstene lamp exposure, wavelength calibration
with a thorium lamp exposure and the removal of cosmic rays. 
Cross-correlation functions are also automatically computed. The background
sky is not automatically subtracted but it can be done later by the observer.  

\begin{table*}[t]
\caption{Veiling values, photospheric, HeI and H$\alpha$ blue and red 
absorption radial velocities, and equivalent widths of the main emission
lines. All radial velocities are expressed in the stellar rest frame
($V_{lsr} = V_{rad}$ - 17.3 \kmsn).}             
\label{spec}                        
\centering                          
\begin{tabular}{lllllllll}          
\hline\hline                        
JD& Veiling & $V_{lsr}$ phot.& $V_{lsr}$ HeI & $V_{lsr}$ (H$\alpha$ blue) & $V_{lsr}$ (H$\alpha$ red) & H$\alpha$ EW & HeI EW & H$\beta$ EW\\
-2450000. & & (km\,s$^{-1}$) & (km\,s$^{-1}$) & (km\,s$^{-1}$) & (km\,s$^{-1}$) & (\AA) & (\AA) & (\AA)\\
\hline                               
3288.82 & 0.17 & 0.8  & 5.4 & -42.5 &  5.2  & 15.6 & 0.3 & 1.7 \\
3289.82 & 0.26 & 0.7  & 4.0 & -22.7 & 32.2  & 12.3 & 0.7 & 1.4 \\
3291.80 & 0.54 & -3.1 & 5.8 & -24.5 & 30.9  & 21.9 & 1.1 & 2.2 \\
3295.80 & 0.15 & -0.3 & 9.7 & -27.2 & 16.0  & 9.5  & 0.2 & 1.3 \\
3296.81 & 0.15 & 0.7  & 9.6 & -37.3 &  4.9  & 15.1 & 0.3 & 2.2 \\
3297.80 & 0.19 & 0.7  & 4.5 & -26.7 & 16.7  & 15.9 & 0.4 & 1.8 \\
3298.84 & 0.22 & 0.9  & 5.3 & -30.9 & 26.9  & 11.8 & 0.7 & 1.2 \\
3308.77 & 0.18 & -0.4 & 8.2 & ---     & ---    & 5.0  & 0.5 & 0.9 \\
3309.81 & 0.31 & -1.0 & 5.2 & ---     & ---    & 8.9  & 0.3 & 0.8 \\
3310.78 & 0.21 & -1.0 & 12.2  & -24.2 & 17.5  & 6.7  & 0.4 & 0.7 \\
3311.76 & 0.14 & -1.0 & 6.1 & -27.7 & 17.2  & 7.3  & 0.2 & 0.9 \\
3312.77 & 0.19 & -0.4 & ---    & -35.0 & 20.9  & 8.4  & 0.2 & ---  \\
3314.75 & 0.15 & 0.5  & 7.2 & ---     & ---    & 63.7 & 0.5 & 4.0 \\
3315.75 & 0.18 & 0.3  & 4.6 & ---     & ---    & 34.0 & 0.5 & 2.8 \\
3320.79 & 0.13 & 0.5  & 11.0  & -33.8 & 22.5  & 13.9 & 0.3 & 0.8 \\
3332.69 & 0.40 & 0.7  & 7.1 & -22.9 & 40.1  & 30.5 & 1.2 & 2.5 \\
3335.71 & 0.24 & -2.0 & 11.7  & -12.9 & 44.4  & 19.5 & 0.5 & 2.0 \\
3337.76 & 0.20 & 0.5  & 7.3 & -23.5 & 48.0  & 23.7 & 0.4 & 2.0 \\
3338.73 & 0.27 & 1.2  & 3.2 & -18.9 & 47.2  & 23.6 & 0.7 & 1.7 \\
3339.73 & 0.41 & 1.7  & 3.8 & -15.9 & 49.6  & 43.1 & 1.4 & 3.4 \\
3340.72 & 0.68 & 3.7  & 7.7 & -12.6 & 52.4  & 44.1 & 1.6 & 2.6 \\
3341.72 & 0.41 & 1.1  & 13.0  & -13.4 & 50.6  & 38.3 & 1.1 & 2.4 \\
\hline                               
\end{tabular}
\end{table*}

\section{Results}
\subsection{Photometry}

\begin{figure}
\includegraphics[scale=.63]{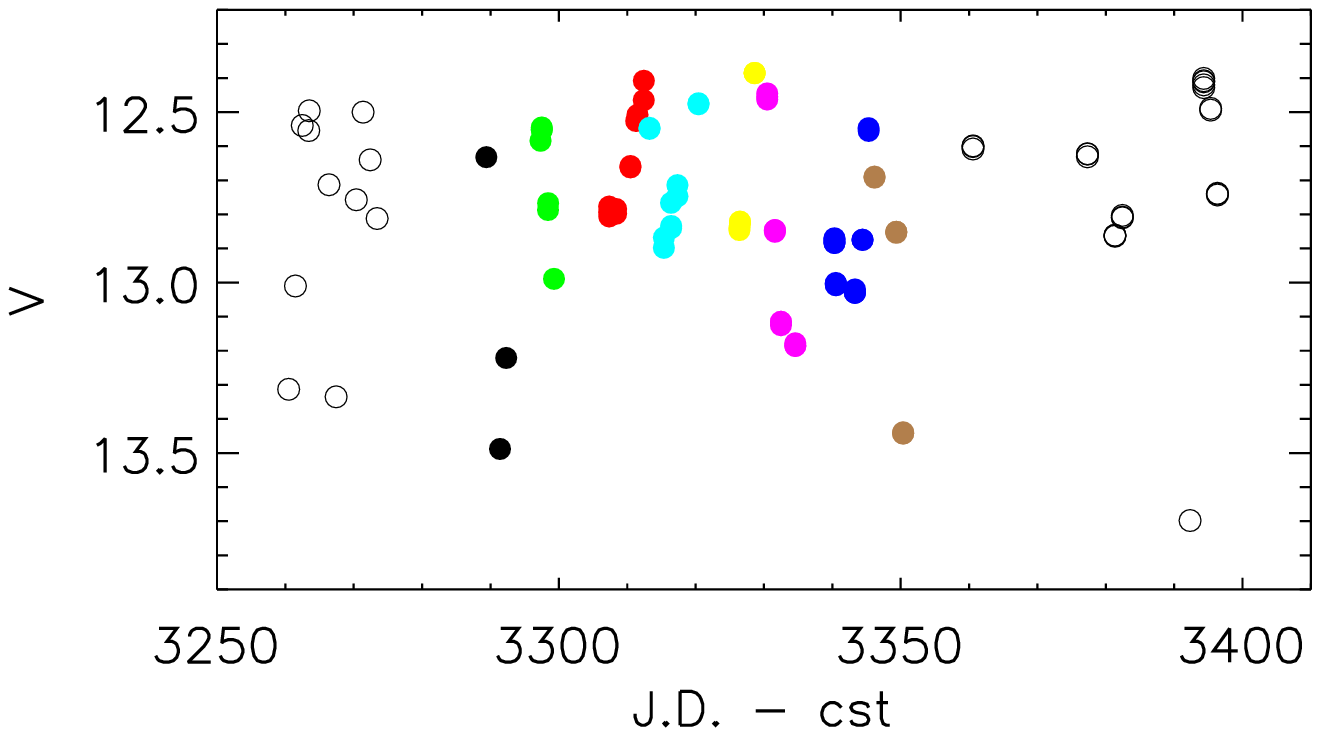}
\includegraphics[scale=.63]{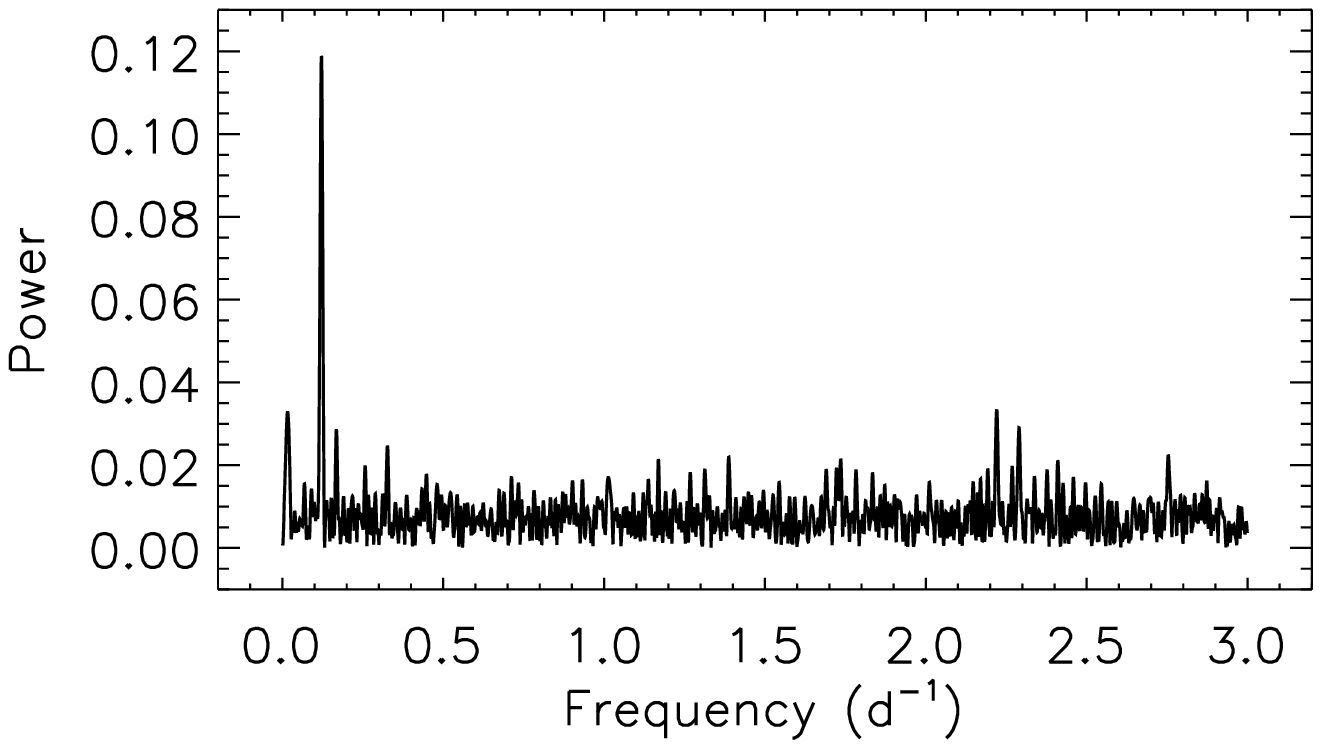}
\caption{The V-band light curve of AA Tau (top panel) and its CLEANed
  periodogram (lower panel). V-band measurements corresponding to
  quasi-simultaneous spectroscopic observations were obtained between
  JD~3285 and 3355 are marked by filled symbols. Different cycles
  are shown by different colors (or shades of grey).} \label{aatauv}
\end{figure}
\begin{figure}
\includegraphics[scale=.63]{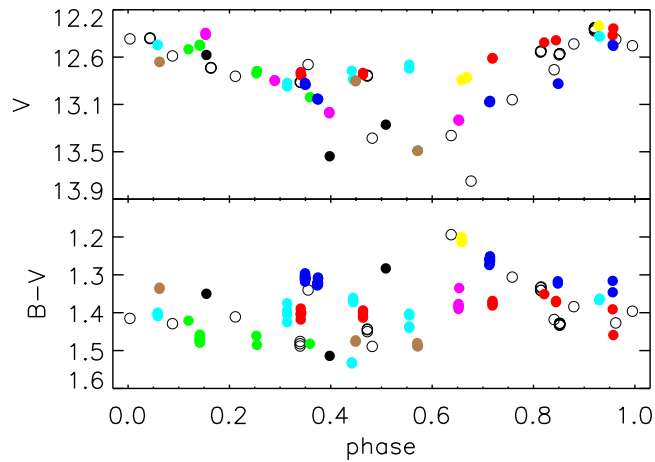}
\caption{AA Tau's V-band (top) and B-V (bottom) light curves folded in
  phase with a period of 8.22 days. Different cycles are shown by
  different colors (or shades of grey).} \label{aatauvp}
\end{figure}

AA Tau's V-band light curve is shown in Figure~\ref{aatauv} over
nearly 150 days. It is characterized by a nearly constant level of
maximum brightness (V$\simeq$12.4) with luminosity drops reaching 1.2
magnitudes. The general photometric behaviour of AA Tau during this
campaign is qualitatively similar to what we had previously reported
for similar light curves obtained in 1995 and 1999 (see Bouvier et
al. 1999, 2003). Periodogram  (Horne \& Baliunas 1986, Roberts et
al. 1987) and string-length (Dworetsky 1983) analyses of the V
and B-band light curves reveal a clear periodic signal with
P=8.22$\pm$0.03 days, consistent with previous estimates (8.2-8.6d,
see Vrba et al. 1989, Shevchenko et al. 1991, Bouvier et al. 1999,
2003). The CLEANed periodogram of the V-band light curve is shown in
the bottom panel of Figure~\ref{aatauv}. The False Alarm
Probability (FAP) was computed following the Monte Carlo method
described in Herbst et al. (2002). The FAP of the highest periodogram
peak was found to be less than 0.01\%. The V-band light curve folded
in phase with this period is shown in Figure~\ref{aatauvp}.
Similar results are obtained for the B-band light curve. The phased
light curve exhibits a broad brightness minimum with an amplitude of
about 1~mag. A large dispersion of the flux is seen to occur within
the deep minimum.

In previous papers of this series, we argued that the 8.22-day period is
  the rotational period of the star. From the modelling of AA Tau's
  spectral energy distribution, Basri \& Bertout (1989) derived an
  inclination angle of 70$\degr$ (see also O'Sullivan et al. 2005). This
  estimate combined with a $v\sin{i}$ of 11.3 \kms (Paper II) and a stellar
  radius of 1.85 R$_\odot$ (Paper I), leads to an expected rotational
  period of 7.8d in rough agreement with the measured photometric
  period. In addition, we report below for the first time a period similar
  to the 8.2-day photometric period for the flux and radial velocity
  variations of emission lines as well as for veiling. This supports the
  8.2-day period as being the stellar rotation period. 

A closer examination of the phased light curve reveals that the amplitude
of modulation varies from cycles to cycles, reaching about 1.2 mag from
JD~3260 to JD~3300 (5 cycles) and again from JD~3330 to JD~3350 and perhaps
later ($\geq$2.5 cycles), but amounting to only 0.3 mag between JD~3305 and
JD~3330 (3 cycles).
Even though the amplitude changes, the phase of the modulation appears to
be conserved over the whole 140 day-long observations.

B-V colours were computed from each quasi-simultaneous B and V
measurements. The phased variations of the B-V colour are shown in
Figure~\ref{aatauvp}. The amplitude of B-V color variations amounts to
0.3~mag only, much shallower than the brightness variations. There is
no clear evidence for a modulation of the B-V color within cycles, and
most of the dispersion results from B-V color changes from one cycle
to the next. It is noteworthy that both the bluest ($\sim$1.3) and the
reddest ($\sim$1.5) B-V colors can occur during the brightness
minimum.

\begin{figure}
\includegraphics[scale=.43]{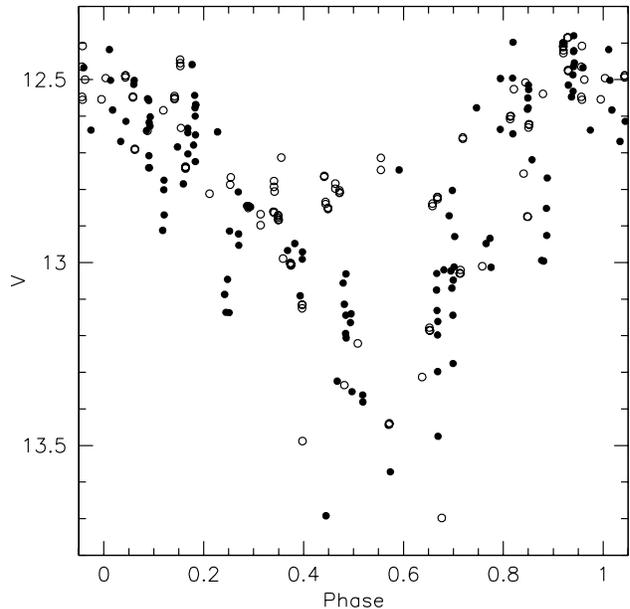}
\caption{AA Tau's phased V-band light curve (P=8.22d) including our
  own measurements (open dots) as well as ASAS ones (filled dots)
  covering over more than 5 months, from JD~3230 to JD~3396. Note the
  phase stability of the light curve on this time span and the range
  of variation in the deep brightness minima.}  \label{asas}
\end{figure}

In addition to the photometry reported here, we found a number of
V-band measurements obtained for AA Tau at the same epoch by the All
Sky Automated Survey (ASAS, Pojmanski et al. 2005). Adding these
measurements to ours, we show the resulting phased light curve in
Figure~\ref{asas}. A periodogram analysis of this dataset
confirms the 8.2-day period at the 0.01\% FAP level. ASAS
measurements also confirm the stability of the phase of the
photometric variations over a timescale of nearly half a year (170
days), while the depth of the photometric minima is much
more variable.

\subsection{Spectroscopy}

\subsubsection{Photospheric lines}

The photospheric lines of CTTS often appear shallower than expected for
their spectral types. This is due to an extra continuum, known as veiling,
produced at the shock region where the infalling material hits the star,
that is added to the stellar continuum.  In order to use the veiling as a
proxy for accretion onto the stellar surface (cf. Paper I), we computed the
veiling using a $\chi^2$ fit, based on the method described by Hartigan et
al. (1989), on six spectral intervals about 50 \AA-wide located between
5400 \AA \ and 6700 \AA. We used as a reference for the veiling
calculations a spectrum of the K7 weak T Tauri star V819 Tau obtained with
HARPS with the same observing setting as AA Tau. This weak T Tauri star was
shown in Paper II to be a very good match to the AA Tau photosphere. The
spectrum of V819 Tau was rotationally broadened to the AA Tau $v\sin{i}$
value determined in Paper II (11.3 \kmsn). The veiling standard
deviation was computed for each night using the veiling values obtained in
the 6 separate orders. The rms deviation ranges from 0.04 to 0.14, with a
mean value of 0.07 over the 22 nights.

\begin{figure}
\includegraphics[scale=.63]{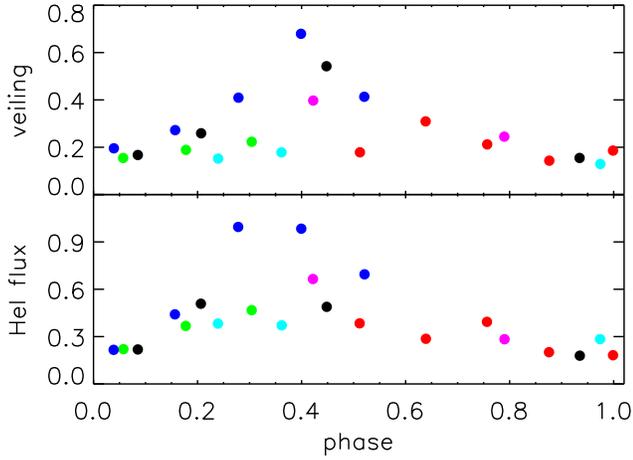}
\caption{Veiling (top) and He I line flux (bottom) variations folded in
  phase with a period of 8.22 days. The origin of phase is the same as
  for photometric measurements. Different cycles are shown by
  different colors (or shades of grey).} \label{veiling}
\end{figure}

\begin{figure}
\includegraphics[scale=.63]{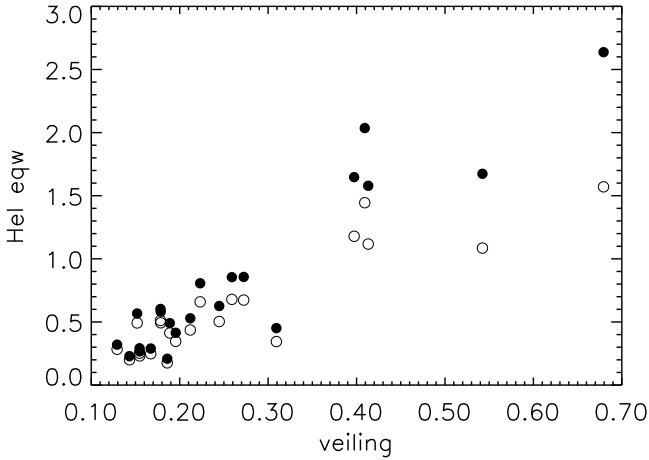}
\caption{ HeI equivalent width plotted against veiling. Filled
  (resp. empty) dots correspond to EW(HeI) measurements on veiled
  (resp., deveiled) spectra.} \label{heveil}
\end{figure}

The derived veiling values are low, ranging from 0.2 to 0.7, but are
considerably higher than in Paper II, where they were all below 0.3.  A
periodogram analysis suggests that the veiling varies periodically with a
8.2$\pm$0.3 day period (FAP=0.22), which is consistent with the stellar
rotation period given the much poorer time sampling of the spectroscopic
data (see Figure~\ref{veiling}). The veiling is weak and almost constant in
the photometric cycles where only a shallow brightness minimum is present,
while it shows a strong peak near phase 0.4 in the deep minimum cycles.  In
Paper II, the veiling presented two maxima per photometric cycle, as if we
were seeing 2 hot spots at the surface of the star, while in 2005 only one
maximum per cycle is visible. We measured the equivalent width of the He I
line by direct integration on the spectrum and computed line fluxes with
the photometric measurements as $F({\rm HeI}) = cst \times EW({\rm HeI})
\times 10^{-0.4m_V}$, where $m_V$ is interpolated from the light curve on
the time of the spectroscopic observations. The rms error on the normalized
HeI flux is of order of 0.1 (cf. Fig.~\ref{veiling}). We observe a strong
correlation between the veiling and the He I line flux, as well as between
veiling and EW(HeI) measured either on the veiled or on the deveiled
spectra (correlation coefficient of 0.9, cf. Fig.~\ref{heveil}), and a
weaker correlation between the veiling and B-V. The B-V vs. veiling
behavior is however still consistent with the 1999 results, lower B-V
values corresponding to higher veiling values.

\begin{figure}
\includegraphics[scale=.43]{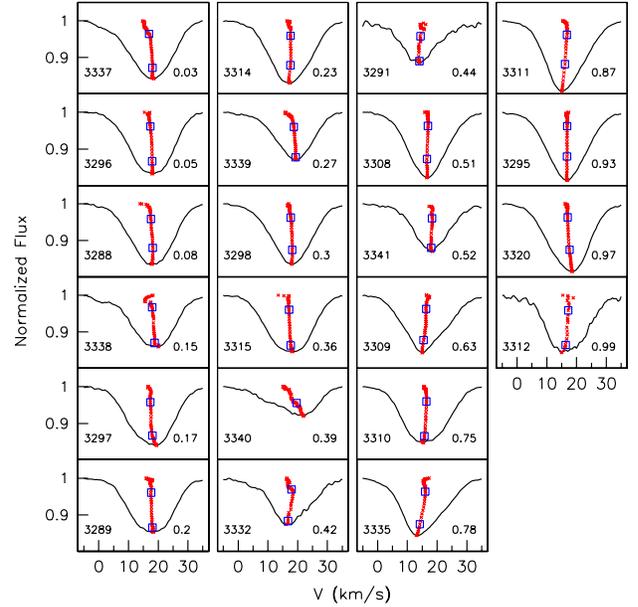}
\caption{Photospheric cross-correlation profiles (CCF). The profiles
  are ordered by increasing phase, as indicated in lower right corner
  of each panel, while Julian date is given in the lower left
  corner. The line bissector (cf. Queloz et al. 2001) is also
  shown (red crosses). The bissector average slope measures the degree
  of asymmetry of the CCF and is estimated as the velocity difference
  between the 2 data points indicated by open squares (see text),
  except for JD~3340 where the CCF is too shallow.}
  \label{ccf}
\end{figure}

Cross-correlation functions (CCFs) of the photospheric spectrum were
automatically computed by the HARPS spectrograph reduction pipeline
using a binary mask corresponding to a K5 template
spectrum. CCFs provide high signal-to-noise information about the
features present in the star's averaged photospheric spectrum. They
were computed for each AA Tau's spectrum and are shown in Figure
\ref{ccf}.  It is seen that the averaged photospheric line profiles
are sometimes asymmetric and this could be the cause of the radial
velocity variation noticed in Paper II. 

We measured the radial velocities adjusting a gaussian to the
cross-correlation profiles in three different ways: fitting the profile
centroid, the far wings, or an intermediate region between the minimum and
the far wings. The mean radial velocity values obtained with the three
different fitting procedures agree with each other ($v_r = 17.3 \pm 1.6$
\kmsn) and are consistent with the value obtained in Paper II ($v_r = 17.1
\pm 0.9$ \kmsn). Individual values can vary by as much as $1.5$ \kms
between different fits, but the three radial velocity data sets presented
the same periodicity of 8.5 $\pm$ 0.4 days when submitted to a periodogram
analysis (FAP=0.02), which is slightly but not significantly different from
the photometric period. All radial velocities below are expressed in the
stellar rest frame ($V_{lsr}$).

\begin{figure}
\includegraphics[scale=.50]{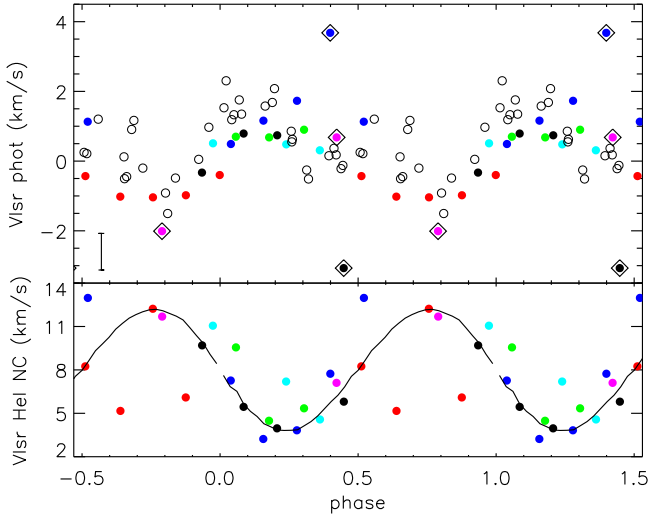}
\caption{{\it Top panel :} Photospheric radial velocities as a
  function of phase. The open circles correspond to the values
  measured in the 1999 campaign (Paper II) and the filled symbols to
  the 2004 campaign. The phase of the 1999 points has been
  arbitrarily scaled to fit that of the 2004 dataset since the phase
  information is lost between the 2 epochs (see text). The points
  with an overplotted diamond correspond to very asymmetric
  cross-correlation profiles.  The error bar in the left corner
  represents the mean error of the 1999 values. {\it Bottom panel :}
  HeI radial velocities. The solid line shows the radial velocity
  variations expected from a hot spot located at a latitude of
  70$\degr$ on the stellar surface.}
\label{vrad}
\end{figure}

In Figure \ref{vrad} we show the radial velocities obtained fitting
the wings of the cross-correlation profiles, overplotted with the
values determined in Paper II.  The radial velocity variations
exhibit the same period and about the same amplitude as those reported
in Paper II. We cannot tell whether the phase is conserved over the 5
year timescale between the 2 epochs. This time span corresponds to
about 220 periods and phase conservation could be tested only if the
period was derived to an accuracy of $\simeq$0.1\% or
better. Nevertheless, the very good agreement between the two datasets
suggests a similar origin for the observed radial velocity variations
at the 2 epochs.

\subsubsection{Survey of emission line profiles}
A sample of AA Tau line profiles was presented in Paper II and the new
observations show overall the same variety of emission profiles.  For
all lines except \haln, we computed residual profiles by subtracting
the continuum normalized and veiled spectra of V819 Tau from the
continuum normalized spectra of AA Tau. In this process, we used
the veiling value measured between 5425 \AA to 6660 \AA, which
exhibits no clear trend with wavelength. The veiling in the \hbeta
region is probably higher but AA Tau's veiling is in general small and
the photospheric features are reasonably well removed from the \hbeta
region when we use a single veiling value.

\begin{figure}
\center
\includegraphics[scale=.43]{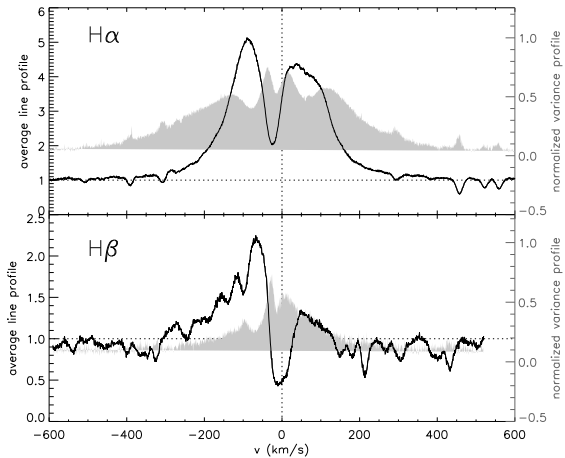}
\caption{Average lines profiles (solid line) and their variance (grey
  shaded area) are shown for \hal (top) and \hbeta (bottom).} \label{var_halpha}
\end{figure}


\begin{figure*}
\center
\includegraphics[scale=.43]{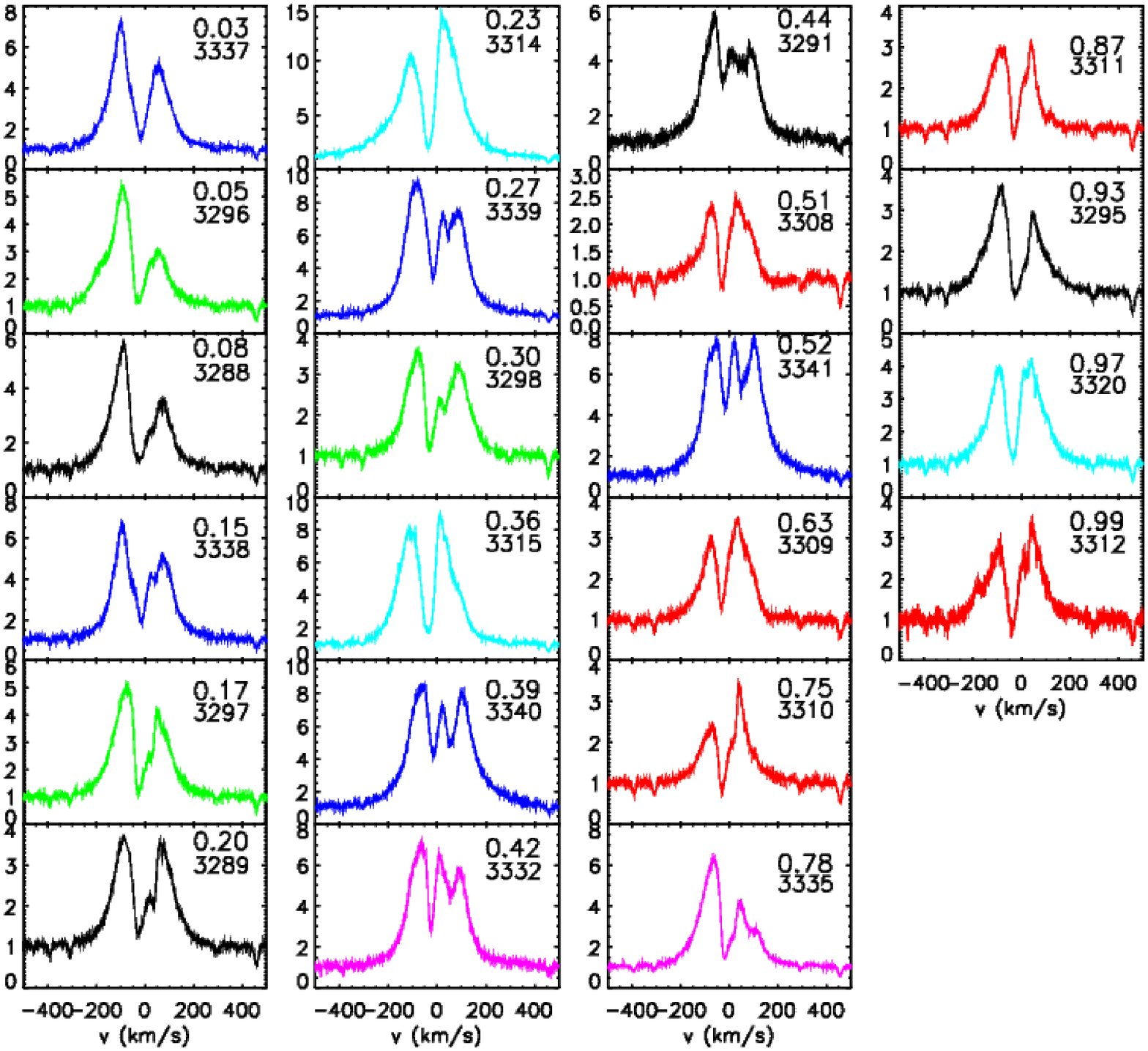}
\caption{\hal emission line profiles ordered by rotational phase (top panel number).
Julian date is given (bottom panel number). Note the appearance of a redshifted 
absorption component from phase 0.39 to 0.52.}\label{halpha_phase}
\end{figure*}

\begin{figure*}
\center
\includegraphics[scale=.43]{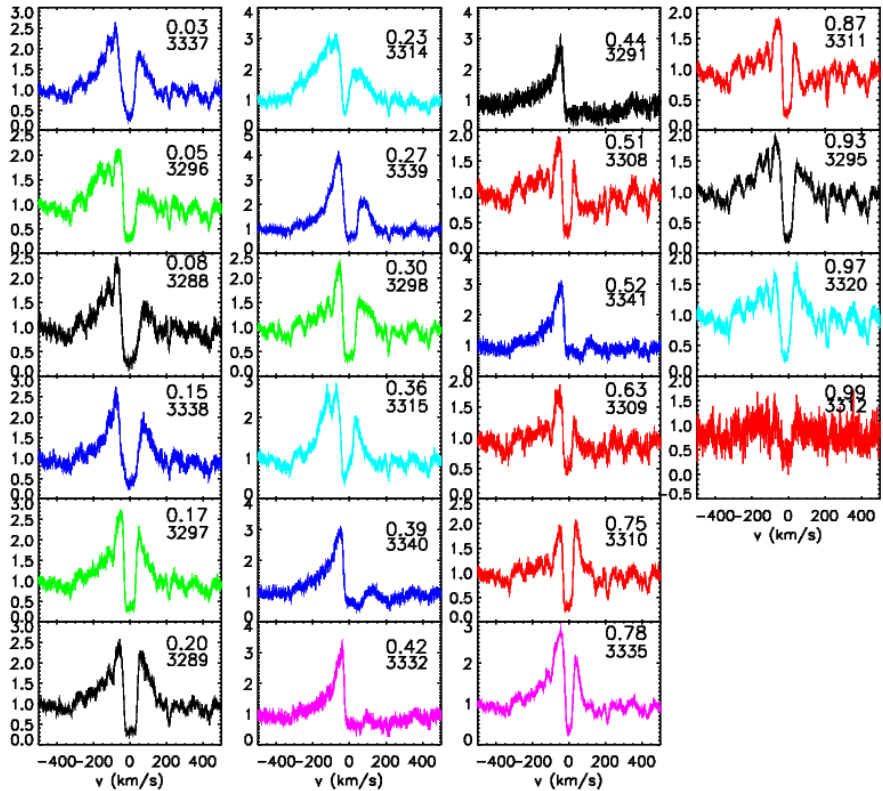}
\caption{\hbeta emission line profiles ordered by rotational phase (top panel number).
Julian date is also given (bottom panel number). Note the appearance of a redshifted 
absorption component from phase 0.39 to 0.52.}\label{hbeta_phase}
\end{figure*}

In Figure \ref{var_halpha} we show the average \hal and \hbeta
profiles as well as their variance (see Johns \& Basri 1995). The
average \hal profile and its variance observed during this new
campaign are quite similar to those reported for the 1999 campaign
(Paper II). This suggests that both the emission line region(s) and
the intrinsic source(s) of variability remained the same over the
years. The \hal average profile is reminiscent of the model \hal
profile computed by Kurosawa et al. (2006) which assumes that the line
is formed partly in an axisymmetric magnetospheric funnel flow and
partly in a slowly accelerating disk wind both seen at a high
inclination (see their Fig.A1). The mean \hbeta profile, with a
triangular blue wing, a central absorption and a secondary peak in the
red wing, closely ressembles the Pa$\beta$ profile computed by
Kurosawa et al. (2005) as arising from an axisymmetric accretion
funnel flow seen at high inclination (cf. their Fig.9). This is
consistent with the high inclination viewing angle to AA Tau
($i\sim$75$\degr$, Paper I). The variance \hal profile also shares
some similarity with variance profiles computed from models in which
the accretion flow is confined to azimuthal curtains instead of being
axisymmetric (Symington et al. 2005b), specifically a primary peak of
variance close to zero velocity and high velocity secondary maxima in
the blue and red wings. We note however that the degree of variability
observed in the red wing of AA Tau's \hal profile is much larger than
predicted by azimuthally confined curtain models.

\begin{figure}
\includegraphics[scale=.57]{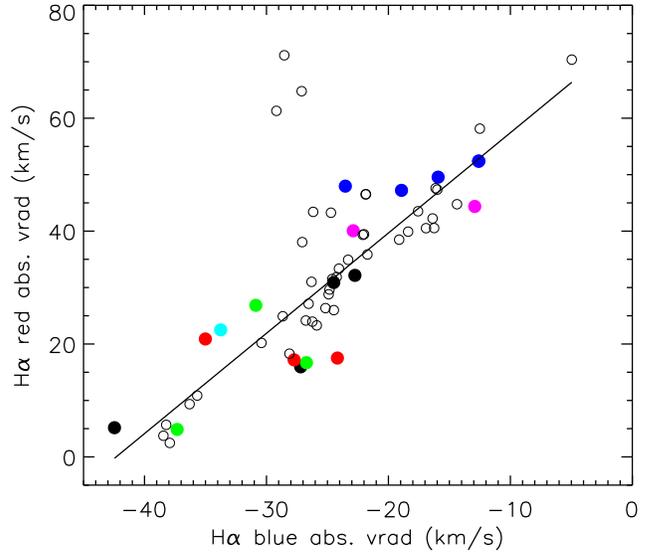}
\caption{Correlation between the radial velocity of the blueshifted (wind) and 
redshifted (accretion) absorption component in the \hal line (color data points: 2004,  
open data points: 1999).}\label{vblue_vred}
\end{figure}

In Figures \ref{halpha_phase} and \ref{hbeta_phase} we show the \hal
and \hbeta emission lines as a function of rotational phase. The deep,
blueshifted central absorption is seen at all times, which presumably
arises from a cold axisymmetric inner disk wind. The broad emission
wings are also always present, which suggests a globally axisymmetric
accretion flow onto the star to first order. However, we notice that
when a deep minimum occurs in the light curve and the photosphere is
obscured by the inner disk warp, a high velocity redshifted absorption
component of \hal is clearly seen, with maximum strength near phase
0.5. This suggests that we are observing downwards along the main
accretion funnel flow at this phase. The redshifted absorption by the
main accretion stream is even more dramatic in \hbeta where the entire
red emission is suppressed at phases near the photometric minimum and
the redshifted absorption is sometimes observed below the continuum
(like JD~3291, 3332, 3340). The cycles that only present a shallow
photometric minimum do not show such a pronounced redshifted
absorption component in the \hal lines (e.g. JD~3308). This is
consistent with the low veiling and low \heI equivalent width values
(Fig.\ref{veiling}) in that case, all of which indicate that accretion
is at a very low level when the obscuring circumstellar material is
absent.

We decomposed the \hal profile using three Gaussians, corresponding to
a centered emission, a blueshifted absorption and a redshifted
absorption.  The decomposition is not always straightforward (see
Fig. 14 of Paper II) but the radial velocities of the three components
are well determined by the Gaussian fits.  We had noticed in Paper II
a tight correlation between the radial velocity of the blueshifted
(wind) and redshifted (accretion) absorption components in the \hal
line that we interpreted as being due to the inflation of the stellar
magnetosphere, caused by differential rotation between the stellar
magnetic field lines and the inner disk.  Such a correlation persists
in the current data and the new values agree very well with the
measurements of Paper II as can be seen in Figure \ref{vblue_vred}.

\begin{figure}
\includegraphics[scale=.50]{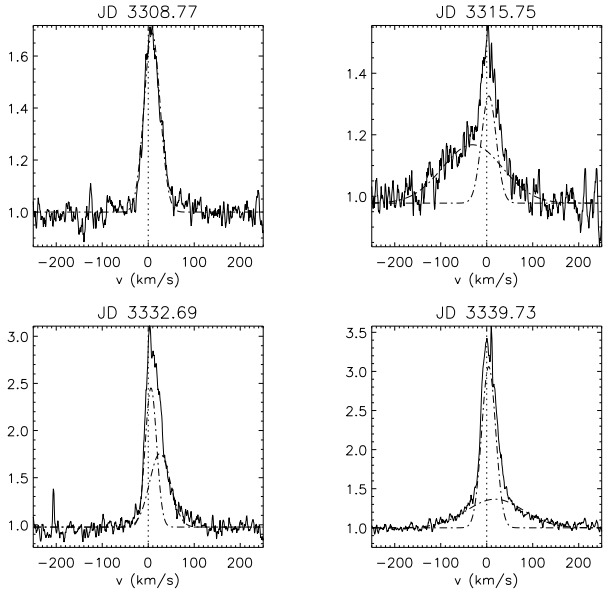}
\caption{\heI line decomposition. The line profile is fitted by either
  one gaussian when only a narrow component is present or two when a
  broad component appears.} \label{decomp_heI}
\end{figure}

The \heI line presents a narrow emission component (NC) that dominates
  the profile and half of the time a broad emission component (BC) as
  well. We then decomposed the \heI line using one or two gaussian
  emissions as shown in Figure \ref{decomp_heI}.  The NC is relatively
  stable kinematically, with radial velocity measurements going from
  $+4$ \kms to $+12$ \kms and a FWHM varying from 16 \kms to 20
  \kmsn. A periodogram analysis of the radial velocity variations
  suggests a possible period of $\sim$8.0$\pm$0.4 days (FAP=0.30),
  consistent with the 8.2-day photometric period
  (cf. Fig.~\ref{vrad}). The equivalent width and the line flux of
  the NC vary considerably in the same period and are well correlated
  with the veiling variations.  According to Beristain et al. (2001)
  the NC is formed near the accretion shock in the post-shock region,
  and should strongly correlate with the veiling, as observed in AA
  Tau.  The BC is much more kinematically variable than the NC, with
  radial velocity values going from $-33$ \kms to $+31$ \kms and FWHM
  ranging from 30 \kms to 40 \kmsn.  It is most of the time redshifted
  but was observed blueshifted twice (JD~3314 and 3315,
  cf. Fig.~\ref{decomp_heI}).  According to Beristain et al. (2001)
  the BC could be either formed in the accretion column (if redshifted
  in excess of 8 \kmsn) or in an accretion-driven hot stellar wind (if
  blueshifted in excess of $-30$ \kmsn). Our results show that a
  single star can move from the accretion dominated BC (redshifted) to
  a wind dominated BC (blueshifted) and back in a short period of time
  (a couple of nights).

\begin{figure}
\includegraphics[scale=.70]{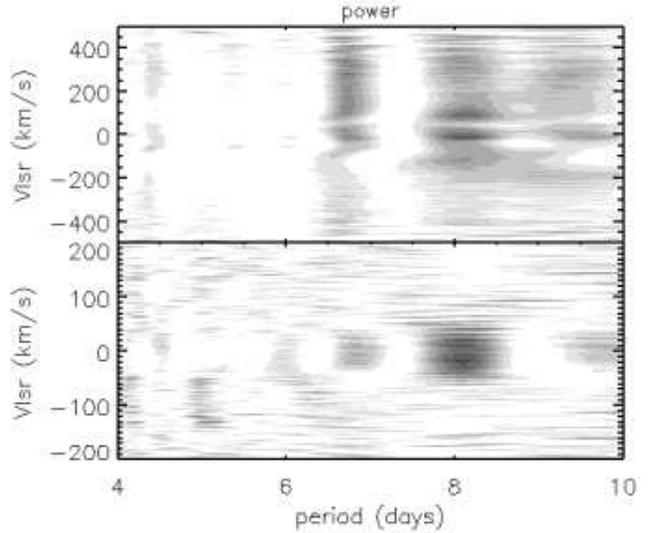}
\caption{Period search results of the \hal (top) and \heI (bottom)
  lines. The power scale ranges from 0 (white) to a maximum
  value of about 6.0 (black).} \label{period}
\end{figure}

In Figure \ref{period} we show our period search result for the \hal
and \heI lines. Periodograms were computed in each velocity
channel on the intensity of the deveiled and normalized line
profiles. Similar results are obtained from the analysis of veiled
profiles. All the main emission lines show periodical intensity
variations around 8.0 $\pm$ 0.4 days, consistent with the more
accurately defined photometric period of 8.2 days.  The maximum power
in the 2D periodogram of the \hal (resp. HeI) line occurs at
$V_{lsr}\simeq$-10 \kms (resp. 3 \kmsn) and corresponds to a FAP of
0.15 (resp. 0.10).  This is the first time we find clear evidence for
the emission line flux being modulated with the same period as the
stellar flux (see Paper II). Most of the emission lines also show more
marginal evidence for periodicities at 6.7 days.

\section{Discussion}

The new dataset obtained on AA Tau provides new and clear evidence for
magnetospheric accretion in this object. We thus discuss in this
section the observed variability pattern of the various diagnostics in
the framework of the magnetically-channelled accretion concept,
starting form the inner disk edge down to the stellar surface. In the
following discussion, the origin of rotational phases ($\Phi = 0$) is
taken as being the epoch of maximum continuum flux in the V-band. This
maximum brightness level has proved very stable over at least the last
10 years (see Papers I, II) and likely corresponds to the unobscured
photosphere. In addition, the veiling is low at this phase, so that
the observed continuum flux is probably close to the star's
photospheric flux (reddened by $A_v$ = 0.8~mag, cf. Paper I). 

\subsection {Characterizing the magnetospheric accretion/ejection region}

As the system rotates with a period of 8.22d, the continuum flux
decreases to reach a minimum around $\Phi\simeq0.5$. This behaviour
had already been reported in previous campaigns, and we argued that
the dimming of the system results from the central star being partly
occulted by circumstellar material (Papers I, II). A supporting
argument comes from the measurement of an increased polarization level
during the eclipses (M\'enard et al. 2003, see also O'Sullivan et
al. 2005). We further argued that the optically thick occulting screen
corresponds to the magnetically-warped, dusty inner disk edge. As the
periodicity of the eclipses is similar to the rotational period
of the star, this locates the inner dust disk edge close
to the corotation radius which, assuming keplerian rotation, is
$r_{co} \simeq 8.8 \ {\rm R}_\star$.

The new dataset confirms both the periodic nature of the eclipses and
their changing amplitude on week timescales (see Paper II). This can
be seen in Fig.~\ref{aatauvp} and ~\ref{asas}. Over nearly 170 days,
corresponding to more than 20 rotations, the phase of the brightness
minimum is conserved. While the depth of most eclipses is of order of
1~mag, we observed shallower eclipses with an amplitude of only a few
0.1~mag during 3 consecutive rotation cycles, located in the middle of
the observing campaign (see Fig.~\ref{aatauv}). We conclude that the
shape of the occulting screen can change abrutply on a timescale of
days, leading to shallower eclipses for a few rotations, and is 
thereafter restored with nearly the same structure. A similar
behaviour had been reported in Paper II for one rotation. 

AA Tau's new light curve exhibits a hint of an asymmetric shape around the
minimum, with a steeper egress than ingress (Fig.~\ref{asas}). At a given
brightness level, the system also appears redder during ingress than during
egress (Fig.~\ref{aatauvp}). The shape of the eclipse reflects the
azimuthal distribution of optically thick dust in the corotating inner disk
warp. Such a slight light curve asymmetry, with a shallow decrease and a
steep rise, may indicate a correspondingly asymmetric azimuthal
distribution of dust at the inner disk edge, as qualitatively predicted by
3D numerical simulations of disk accretion onto an inclined dipole
(Romanova et al. 2003).


If the warp of the inner disk edge results from its interaction with
the stellar magnetosphere, as argued in Paper I, the dominant
accretion funnel flows ought to face the observer close to the middle
of the occultation. The analysis of the Balmer line profiles supports
this view. Fig.~\ref{halpha_phase} and ~\ref{hbeta_phase} clearly
reveal the development of high velocity redshifted components in the
profiles of both lines (also seen in H$\gamma$) from $\Phi=0.39$ to
$\Phi=0.63$, i.e., around the center of the eclipse. High velocity
redshifted components are expected to be seen when the observer looks
downwards the accretion funnel flow against the accretion shock
(Hartmann et al. 1994). The line profiles thus indicate that the
center of the eclipse also corresponds to the time at which the
accretion shock faces the observer. This strongly suggests that the
inner disk warp is magnetically-driven indeed (Terquem \& Papaloizou
2000, Lai 1999).

The line profiles observed during one of the shallower eclipses
(JD~3308, $\Phi=0.51$) exhibit a much shallower redshifted component
than the profiles observed at similar phases during deep minima
(e.g. JD~3341, $\Phi=0.52$, see Fig.~\ref{halpha_phase} and
~\ref{hbeta_phase}). As the veiling was also measured to be weaker
during shallow eclipses (see Fig.~\ref{veiling}), this suggests that
the accretion rate onto the star was smaller at that time.  Hence,
there seems to be a clear link between the structure of the occulting
warp at the inner disk edge and the accretion rate onto the star, both
presumably depending on the time variable magnetic configuration at
the disk truncation radius. In addition, {\it blueshifted} HeI broad
emission components were observed only during one of the shallower
eclipses (JD~3314, 3315), which suggests that a hot wind develops as
accretion is depressed onto the star. Finally, the H$\beta$ line
profile reveals transient but significant redshited absorption
components at phases opposite to the deep minima (e.g. JD~3295,
$\Phi=0.05$). We speculate that these transient absorptions may be
related to the viewing of funnel flows against the accretion shock
located at the opposite magnetic pole on the stellar surface
($\Phi\simeq0$, see Fig.10 in Paper I).

The visibility of the hot spot at the stellar surface is measured by
the variations of the HeI line flux originating in the accretion shock
at the bottom of the accretion funnel (Beristain et al. 2001). The HeI
line flux, usually dominated by a narrow component
(Fig~\ref{decomp_heI}), is modulated with the same 8.2d period as the
other diagnostics (Fig.~\ref{period}) and shows a low level modulation
of about a factor of 2, with a maximum around $\Phi=0.4$
(Fig.~\ref{veiling}). The veiling is modulated in the same way and
also peaks slightly before the center of the eclipse
(Fig.~\ref{veiling}).  The close temporal coincidence between the
maximum intensity of accretion shock diagnostics ($\Phi\simeq
0.3-0.6$), the appearance of high velocity redshifted absorptions in
Balmer line profiles ($\Phi=0.39-0.63$), and the occurence of
photometric eclipses ($\Phi\simeq 0.3-0.8$) strongly supports the view
that material is being preferentially accreted from the warped inner
disk edge along magnetic field lines down to the stellar surface.

In addition to the low level modulation, both the HeI line flux and
the veiling exhibit intrinsic variations of larger amplitude on a
shorter timescale. This is examplified by the 4 highest He I flux
values in Fig.~\ref{veiling} and the corresponding measurements of
veiling in Fig.~\ref{veiling}. These episodes occur during the
eclipses ($\Phi=0.3-0.5)$. Whether these are transient episodes on a
timescale of a few hours, such as accretion bursts or opacity changes
in the occultation screen as previously reported in Paper I, or a
slightly varying accretion rate onto the star from one cycle to the
next cannot be easily decided from our dataset.

\subsection{The origin of periodic radial velocity variations}

One puzzling aspect of AA Tau's behaviour is the periodic radial
velocity variations of the photospheric spectrum. We had reported this
result in Paper II and offered possible interpretations, such as a
planetary mass companion orbiting the star or the presence of a large
cold spot at the stellar surface, none of which seemed
satisfactory. The new dataset provides additional evidence for the
modulation of the radial velocity of the photospheric spectrum with a
period of 8.2d and an amplitude of a few km.s$^{-1}$, similar to what
we previously reported (see Fig.~\ref{vrad}). The source of radial
velocity variations thus appears to be the same at the 2
epochs. The coincidence between the period of the radial velocity
variations and the rotational period of the star strongly suggests the
former results from spot modulation. Also, the amplitude of the radial
velocity variations ($\sim$3 \kmsn) is a fraction of the star's
rotational velocity ($v\sin{i}$=11.3 \kmsn), consistent with spot
modulation.

The high resolution spectral series obtained during the new campaign
supports this view. The extrema of the radial velocity curve occur
around rotational phases 0.2-0.3 (maximum) and 0.7-0.8 (minimum),
while the star's rest velocity is measured around phases 0.0 and 0.5,
i.e. at maximum and minimum brightness (cf. Fig~\ref{vrad}).  We
argued above from the line profile analysis that the hot accretion
shock faces the observer around $\Phi=0.5$. Assuming the hot spot is
responsible for the modulation of the photospheric line profiles, one
would expect to measure the star's rest velocity when the spot faces
the observer ($\Phi=0.5$) as well as at the opposite phase ($\Phi=0$),
while the largest $V_{rad}$ excursions ought to occur when the hot
spot is close to the stellar limb ($\Phi\simeq 0.25$ and $0.75$). This
is precisely what the phased radial velocity curve indicates, with
$V_{lsr}\sim$ +1 \kms and -1 \kms at $\Phi\simeq$ 0.25 and 0.75,
respectively. In addition, the radial velocity of the HeI narrow
component (NC) also appears to be modulated, though with a larger
scatter. It varies between $\sim$ +4 \kms and $\sim$ +12 \kms around
$\Phi=0.25$ and $\Phi=0.75$, respectively, i.e. with a phase opposite
to that of the photospheric radial velocity variations, as expected
from hot spot modulation. The time averaged velocity of the HeI line
NC ($V_{lsr}\sim$ 8 \kmsn) presumably reflects the postshock velocity
at the magnetospheric footpoint on the stellar surface (Beristain et
al. 2001).

Fig~\ref{vrad} also shows that a few $V_{rad}$ measurements strongly
depart from the low level periodic $V_{rad}$ variations. These
measurements where obtained during photometric minima ($\Phi=0.39,
0.42, 0.44$; JD~3340, 3332, 3291) at a time when the veiling was also
the highest. The corresponding correlation profiles (CCFs) are
extremely asymmetric (cf. Fig.~\ref{ccf}) and such asymmetries may
conceivably alter the accuracy of the $V_{rad}$ measurements. CCFs
obtained outside theses phases show some level of asymmetry as well,
but usually not as pronounced, as expected from the modulation by a
hot surface spot (e.g. Petrov et al. 2001). The largest CCF
asymmetries observed during the deep photometric minima suggest that
the extreme $V_{rad}$ excursions are related to the partial
occultation of the central star by the asymmetric inner disk warp, and
are superimposed onto the lower level $V_{rad}$ modulation by the hot
spot. Note that only few measurements strongly depart from the hot
spot $V_{rad}$ modulation, which suggests that the azimuthal structure
of the inner disk warp occulting the star is relatively smooth, as
also suggested by the shape of the photometric eclipses.

Hence, the modulation of the photospheric radial velocity in AA Tau's most
likely results from an inhomogeneous brightness distribution on the stellar
surface. Its phase is consistent both with the visibility of the hot
accretion spot derived above from veiling and HeI flux measurements, and
with the HeI line radial velocity variations. For a hot spot occupying a
few percent of the stellar photosphere, the radial velocity amplitude of
the HeI line is related to the latitude $\theta$ of the accretion shock at
the stellar surface by : $\Delta V_{rad} = 2\ v\sin{i} \cdot
\cos\theta$. With $\Delta V_{rad} \simeq 8$ \kms and $v\sin{i} = 11.3$
\kmsn, we derive $\theta \simeq 70 \degr$. The solid line in Fig~\ref{vrad}
shows the expected HeI radial velocity variations from a model featuring a
small hot spot located at a latitude of 70$\degr$. The model curve fits
reasonably well the observed HeI radial velocity variations, at least when
deep eclipses are seen in the light curve. As the accretion shock is
thought to occur at the feet of the funnel flows on the stellar surface
close to the magnetic pole, we deduce that the axis of the large scale
magnetosphere is tilted relative to the star's spin axis by about
20$\degr$. This is consistent with the 12$\degr$ tilt derived by Valenti \&
Johns-Krull (2004) from spectropolarimetric measurements, assuming a
stellar inclination of 66$\degr$. Note that the radial velocity of the HeI
line is modulated around a mean value of $V_{lsr}\sim$ 8 \kmsn, the
putative post-shock velocity, and not around the photospheric velocity, as
would be expected for a small single hot spot. This suggests a more complex
structure for the accretion shock, such as a circumpolar ring around the
magnetic axis (Mahdavi \& Kenyon 1998, Romanova et al. 2004).

The observed periodic radial velocity variations of absorption
and emission lines can thus be, at least partly, accounted for by the
presence of a hot spot at the stellar surface. Whether a hot spot {\it
alone} fully accounts for the radial velocity curve of the
photospheric spectrum is however unlikely. Firstly, we note that the
amplitude of the $V_{rad}$ variations is about the same in 1999 and in
2003 (cf. Fig.~\ref{vrad}) while the veiling was significantly higher
at the latter epoch. If hots spot were responsible for the radial
velocity excursions, one might expect the amplitude of the $V_{rad}$
curve to increase with the spots brightness, which is not seen
here. Secondly, the line shape distorsion caused by a hot continuum
spot is the same as that procuded by a cold spot (e.g. Johns-Krull \&
Hatzes 1997). In Paper II, we found that a cool spot would have to
cover about 50\% of the stellar surface to produce the observed
$V_{rad}$ variations. This is much larger than the expected size of AA
Tau's accretion spots, of order of a few percent (Paper I). Hence, it
is unlikely that a hot spot alone can account for the observed
$V_{rad}$ amplitude. Additional sources of variations are probably
required, such as the occulting effect of the warp inner disk and/or
large cold spots clustered around the hot accretion spot at
the stellar surface. We also note that similar radial velocity
variations were found in the more active CTTS RW Aur and interpreted
as resulting from the modulation by large chromospheric spots (Petrov
et al. 2001). Detailed Doppler Imaging of the star would be needed to
address this issue. Note however that these additional components do
not affect the conclusion regarding the radial velocity variations of
the HeI NC, which more directly traces the accretion hot spot on the
stellar surface. 

\begin{figure}
\includegraphics[scale=.43]{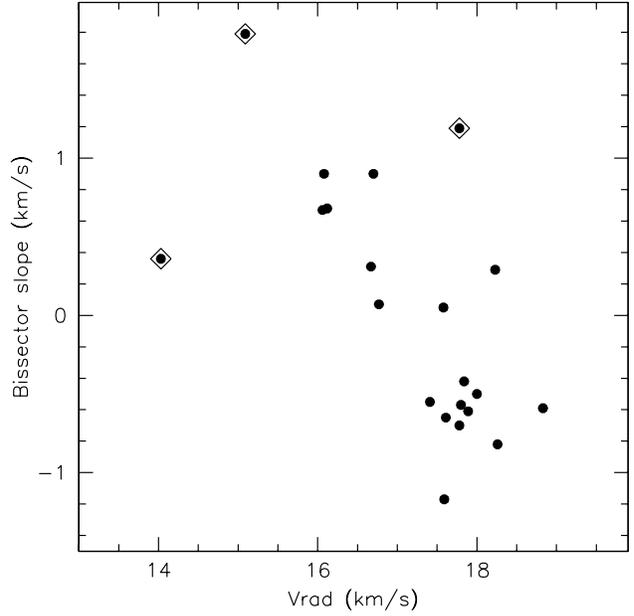}
\caption{The CCF bissector ``slope'' is plotted as a function of the
  measured photospheric radial velocity. The points with an
  overplotted diamond correspond to very asymmetric CCFs.  The inverse correlation seen
  between the 2 quantities indicates that radial velocity variations
  mostly result from spots at the stellar surface.}  \label{biss}
\end{figure}

A final indication that the V$_{rad}$ modulation primarily
  results from stellar spots is provided by the analysis of the line
  profile shape and its deformation as a function of rotational
  phase. Following the method outlined by Queloz et al. (2001), we
  computed the bissector of the photospheric cross-correlation
  function (CCF) at each phase (cf. Fig.~\ref{ccf}). The bissector
  slope is seen to vary as a function of phase. The average slope is
  computed as the velocity difference between the upper and lower
  regions of the bissector, $V_{u}$-$V_{l}$, where $V_{u}$
  (resp. $V_{l}$) is the median velocity of the bissector in the
  intensity range 0.93-0.98 (resp. 0.80-0.85). The bissector slope is
  plotted as a function of the measured V$_{rad}$ in
  Fig.~\ref{biss}. A clear inverse correlation is seen between the
  bissector slope and $V_{rad}$, as expected for spot-induced
  $V_{rad}$ variations (Queloz et al. 2001). In contrast, no such
  correlation would exist if the periodic $V_{rad}$ variations were
  due to stellar oscillations or to the reflex motion of the star in
  response to an orbiting body (Dall et al. 2006). A few measurements
  strongly depart from the observed correlation, as already noticed
  above for $V_{rad}$ measurements (cf. Fig.~\ref{vrad}), and may
  indicate the additional effect of the inner disk warp partly
  occulting the stellar photosphere at times of deep photometric
  minima. 

\subsection {Non steady accretion on a week timescale} 

On a timescale longer than the star's rotational period, we confirm
the correlation we reported in Paper II between the radial velocity of
the blueshifted (wind) and redshifted (accretion) absorption
components of the \hal profile (see Fig.~\ref{vblue_vred}). In Paper
II, we interpreted this correlation as ``magnetospheric inflation
cycles''. Following dynamical models of magnetic star-disk interaction
(e.g. Goodson \& Winglee 1999) accretion from the inner disk onto the star
leads to the inflation of the magnetic funnel flows due to
differential rotation between the inner disk edge and the stellar
surface. Past some critical point, the field lines open and reconnect,
then reducing the accretion flow onto the star and simultaneously
leading to enhanced wind outflow. After field lines have reconnected,
the initial magnetic dipolar configuration is restored and a new cycle
starts. We argued in Paper II that the projected radial velocity of
the redshifted absorption components of the \hal line measure the {\it
curvature} of the accretion funnel flow : as the magnetosphere
inflates, the projected radial velocity of the redshifted component
decreases (see Fig.19 in Paper II). If the collimated wind originates
at or close to the disk inner edge, the projected radial velocity of
the blueshifted absorption component simultaneously increases, thus
yielding the observed correlation. Note that this simple geometric
interpretation favors inner disk winds (e.g. Alencar et al. 2005) over
accretion driven stellar winds (e.g. Matt \& Pudritz 2005).

The results of the new campaign tend to confirm this interpretation.
Just before and during the episode of shallow eclipses and depressed
accretion onto the star (JD~3305-3330), the projected radial velocity
of the accretion flow is small (from +5 to +30 \kmsn) and that of the
inner disk wind is large (from -45 to -25 \kms). According to the
above interpretation, this would correspond to a phase of field line
inflation and opening. The lower accretion rate onto the star is also
accompanied by the development of a hot wind outflow as traced by the
appearence of a broad blueshifted component in the HeI profile. Just
after this episode, as deep eclipses and enhanced accretion resume,
the projected radial velocity of the accretion flow is larger (from
+40 to +50 \kmsn) and that of the inner disk wind is smaller (from -25
to -10 \kmsn), as expected if the initial dipolar magnetospheric
configuration has been restored at this point. The observed timescale
for the disappearance and restoration of deep eclipses is of order of
a week (see Fig.~\ref{aatauv}). The dissipation timescale of the inner
disk warp through bending waves, $\tau_w$, in response to the
disturbance of the underlying magnetic configuration, is given by
$\tau_w = 2 r_c/c_s = 2 r_c/(\Omega\cdot h)$, where $r_c$ is the disk
truncation radius, $c_s$ is the local sound speed, $\Omega$ the
keplerian velocity, and $h$ the warp thickness (Terquem,
priv. comm.). At a distance $r_c=8.8$ R$_\star$ and for a warp
vertical thickness $h/r_c\sim0.3$ (cf. Paper I), this yields $\tau_w
\sim$ 4 days, which is consistent with the observations.

The variations of the accretion and outflow diagnostics on a timescale
of several rotation periods in AA Tau can thus be consistently
understood as resulting from episodes of strong accretion onto the
star together with reduced or absent hot wind (whenever the dipolar
magnetosphere allows the disk material to fall onto the star)
alternating with episodes of reduced accretion and enhanced hot wind
possibly driven by magnetic reconnections (when the magnetosphere
inflates due to differential rotation), with the whole process
occurring on a timescale of a few weeks. In addition, the stability of
the deep central absorption in the Balmer line profile suggests that a
cold inner disk wind is always present. These results tend to support
the concept of magnetospheric inflation cycles on a timescale of
several rotation periods in accreting T Tauri stars, although the
periodic character of such cycles as predicted by dynamical models
remains to be established from longer time series.

\begin{figure}
\includegraphics[scale=.43]{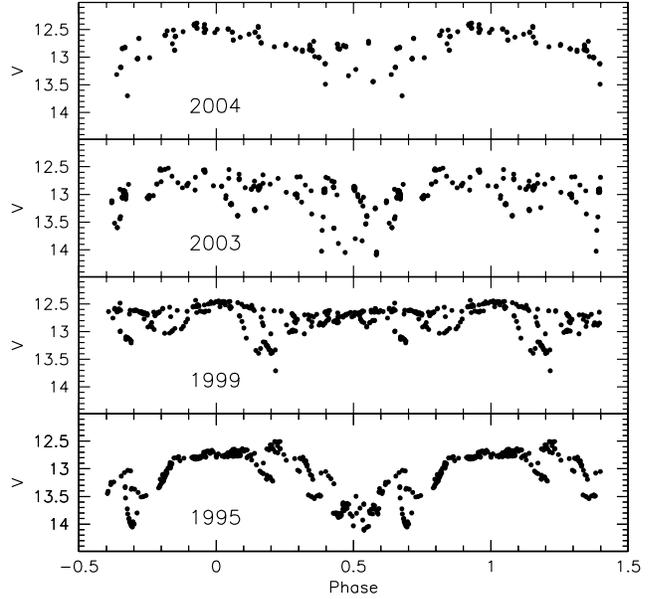}
\caption{V-band light curves of AA Tau at different epochs: 1995
  (Paper I), 1999 (Paper II), 2003 (unpublished), 2004 (this
  paper). All light curves have been folded in phase with the same
  period of 8.22d but the origin of phase for each season is
  arbitrary. Note how the shape and depth of the eclipses vary on a
  timescale of several years.} \label{prevlc}
\end{figure}

\subsection{AA Tau's long term photometric behaviour}

Finally, Fig.~\ref{prevlc} offers a long term view of AA Tau's
photometric behaviour, as recorded in the last 10 years. The
shape and depth of the eclipses do vary on a timescale of weeks within
each observing season, and small phase changes (up to 0.1-0.2) occur
as well on this time scale. Nevertheless, the overall pattern of
photometric variability remains globally stable within each season
over a timescale of several months (Fig.~\ref{asas}).
Fig.~\ref{prevlc} indicates that the long term (years) evolution of
the inner disk structure can be more drastic. The set of light curves
obtained at different epochs (1995, 1999, 2003, 2004) exhibit quite
different patterns. In 1995, a broad and deep eclipse was seen per
rotation. In 1999, 2 shallower eclipses were observed per cycle. The
2003 light curve exhibits a clear primary eclipse, whose amplitude
varies, and may still bear evidence for a shallow secondary
eclipse. The 2004 light curve reported here displays only one eclipse
per cycle, broader and shallower than observed in 1995. The changing
pattern of the eclipses over the years reflects the varying structure
of the inner disk warp occulting the star. This suggests that the
large scale component of the stellar magnetic field interacting with
the inner disk and producing the warp can significantly vary over the
years.

\section{Conclusions}

We have analyzed a long time series of high resolution spectra and
simultaneous photometry obtained for the accreting T Tauri star AA
Tau. The photometric diagnostics provide clues to the structure of the
dusty inner disk edge interacting with the star's magnetosphere while
various spectral diagnostics probe the accretion funnel flows onto the
star as well as the accretion shock at the stellar surface. To first
order, the observed variability of these diagnostics on a timescale
ranging from days to months supports the view of time-dependent,
magnetically-controlled accretion from the inner disk edge onto the
star.

The stellar magnetosphere disrupts the inner disk at the corotation
radius, located about 9R$_\star$ above the photosphere. At the disk
truncation radius, material is lifted away from the disk plane by the
inclined stellar magnetosphere which results in an asymmetric dusty
warp at the disk inner edge. The inner disk warp corotates with the
star and partly occults the stellar photosphere periodically
(P=8.22d), yielding the deep and broad periodic eclipses observed in
the light curve. As the star is periodically occulted, accretion
diagnostics are the strongest (veiling, HeI flux), which indicates
that the accretion shock at the stellar surface is located at the same
rotational phase as the inner disk warp. High velocity redshifted
absorption components also appear in the Balmer line profiles at this
rotational phase, being formed in the main accretion flow which
connects the inner disk to the accretion shock. Hence, from the
temporal coincidence of various photospheric and spectral diagnostics,
we are able to demonstrate for the first time the spatial continuity
between the inner disk warp, the main accretion funnel flow, and the
accretion shock at the stellar surface, as qualitatively expected from
models in which accretion from the inner disk onto the star is
mediated by a large-scale inclined magnetosphere. The line variability
is thus modulated by the corotation of the asymmetric magnetospheric
funnel flow onto the star. Nevertheless, some components of the
emission line profiles remains fairly stable, which suggests that the
degree of asymmetry of the magnetospheric structure and of the inner
disk wind is moderate. By modelling the radial velocity variations of
the star and of the accretion shock, we derive an inclination of
20$\degr$ between the axis of the large scale magnetosphere and the
stellar rotational axis.

On a timescale of several rotation periods, we also find that the
accretion rate onto the star varies, with episodes of strong accretion
and reduced hot wind alternating with episodes of weaker accretion and
stronger hot wind. As accretion onto the star weakens, the depth of
the photometric eclipses is also significantly reduced. This is
readily explained by the fact that the accretion flow onto the star
and the disk inner warp are both driven by the underlying magnetic
structure interacting with the disk. In other words, time-variable
eclipses and accretion rate onto the star both reflect temporal
variations in the topology of the magnetic field lines which connect
the star to the inner disk. We find that the observed variability of
accretion and wind diagnostics in the Balmer and HeI line profiles on
a timescale of weeks is consistent with ``magnetospheric inflation'',
i.e., field line expansion resulting from differential rotation
between the inner disk and the stellar surface. As the magnetic field
lines inflate and eventually become open, the amount of disk material
that can be loaded into closed field lines decreases, thus reducing
the accretion rate onto the star, while more material can be driven
away along open field lines, thus leading to an enhanced hot wind,
possibly driven by magnetic reconnections close to the star. In
addition, evidence for a cold inner disk wind is seen at all
times. The timescale for the development of magnetospheric inflation
is several (typically $\sim$5-10) rotational periods.

Finally, the comparison of AA Tau's light curves obtained over a
timescale of 10 years, between 1995 and 2004, indicates that, even
though periodic eclipses are seen at all epochs on a rotational
timescale, their shape and depth drastically varies on a timescale of
years. This indicates that, in addition to transient perturbations of
the magnetospheric structure occuring on a timescale of weeks
(``magnetospheric inflation''), the overall magnetospheric structure
does change quite substantially over the years. This large-scale
changes may conceivably be related to long term variations in the disk
accretion rate or be intrinsic to the stellar magnetic field such as,
e.g., magnetic cycles.

\begin{acknowledgements}
We thank A. Pal, J. Benko, S. Csizmadia, Z. Kiss, A. Kospal, M. Racz,
K. Sarneczky, and R. Szabo who took part in the observations obtained
at Konkoly Observatory. We thank the referee, C. Johns-Krull, for
insightful comments on the manuscript. SHPA acknowledges support
from CNPq (grant 201228/2004-1), Fapemig and LAOG. Z.~Balog received
support from Hungarian OTKA Grants TS049872, T042509 and T049082.
\end{acknowledgements}

\end{document}